\documentclass[a4paper,11pt]{article}

\pdfoutput=1

\usepackage{jcappub_ver}
\usepackage{amsmath}
\usepackage{graphicx}

\usepackage[english]{babel}
\usepackage[utf8]{inputenc}
\usepackage{pdflscape}
\usepackage{enumerate}
\usepackage{amsbsy}
\usepackage{amsmath} 
\usepackage{graphics}
\usepackage{mathrsfs}
\usepackage{wrapfig}
\usepackage{mathtools}

\usepackage{physics}
\usepackage{breqn}

\usepackage{amsfonts}
\usepackage{pstricks}
\usepackage{color}
\usepackage{setspace}

\newcommand{\bea}{\begin{eqnarray}} \newcommand{\eea}{\end{eqnarray}}
\newcommand{\el}{\nonumber \\}
\newcommand{\re}[1]{(\ref{#1})}

\newcommand{\pat}{\partial}
\renewcommand{\sec}[1]{section \ref{#1}}

\newcommand{\para}{\paragraph}

\renewcommand{\a}{\alpha}
\renewcommand{\b}{\beta}
\renewcommand{\c}{\gamma}
\renewcommand{\d}{\delta}

\newcommand{\ha}{\frac{1}{2}}

\newcommand{\rmd}{\mathrm{d}}
\newcommand{\diag}{\mathrm{diag}}

\newcommand{\ie}{i.e.\ }
\newcommand{\eg}{e.g.\ }

\newcommand{\adot}{\dot{a}}

\title{Higgs inflation and teleparallel gravity}

\author[a]{Sami Raatikainen}
\author[a,b]{and Syksy R\"{a}s\"{a}nen}

\affiliation[a]{University of Helsinki, Department of Physics and Helsinki Institute of Physics,\\ P.O. Box 64, FIN-00014 University of Helsinki, Finland}

\affiliation[b]{Birzeit University, Department of Physics \\
P.O. Box 14, Birzeit, West Bank, Palestine}

\emailAdd{sami.raatikainen@helsinki.fi}
\emailAdd{syksy.rasanen@iki.fi}

\abstract{
Teleparallel gravity is a formulation of general relativity that is physically equivalent to metric gravity if the gravitational action has the Einstein--Hilbert form and matter is minimally coupled. However, scalar fields generally couple directly to the connection, breaking the equivalence. In particular, this happens for the Standard Model Higgs. We show that a teleparallel theory with a non-minimally coupled scalar field has no linear scalar perturbations, and therefore cannot give successful inflation, unless the non-minimal coupling functions satisfy a particular relation. If the relation is satisfied, Higgs inflation can give an arbitrarily large tensor-to-scalar ratio $r$. Our results also apply to $f(T)$ theories, as they are scalar-tensor theories written in different field coordinates. We discuss generalisation to more complicated actions.
}

\begin{document}

\begin{flushleft}
	\hfill		 HIP-2019-34/TH \\
\end{flushleft}

\maketitle
  
\setcounter{tocdepth}{2}

\setcounter{secnumdepth}{3}

\section{Introduction} \label{sec:intro}

\para{Alternative formulations of general relativity.}

The theory of general relativity can be formulated in a number of ways that are conceptually distinct but mathematically equivalent, at least at the classical level and with the simplest action. Among the most well known are three formulations where the connection is specified in different ways a priori: the metric formulation (also known as the Hilbert formulation), the teleparallel formulation \cite{Einstein:1925, Einstein:1928a, Einstein:1928b, Einstein:1930, Unzicker:2005, Aldrovandi:2013, Golovnev:2018red, Krssak:2018ywd} and the symmetric teleparallel formulation (also called coincident gravity and purified gravity) \cite{Nester:1998, Adak:2005, Adak:2006rx, Adak:2008gd, BeltranJimenez:2017, Jarv:2018}. Each of these variants is based on one of the three tensors that completely characterise the geometry of a spacetime manifold: Riemann curvature $R^\alpha{}_{\beta\gamma\delta}$, torsion $T^\c{}_{\a\b}=2\Gamma^{\c}_{[\a\b]}$ and non-metricity $Q_{\c\a\b}=\nabla_\c g_{\a\b}$. These tensors correspond to parts of the connection $\Gamma^\c_{\a\b}$ that encode the three different possible changes from parallel transporting a vector: change in orientation, origin and norm. In the metric formulation, gravity is contained in the Riemann tensor, and torsion and non-metricity are put to zero by imposing the Levi--Civita connection. In the teleparallel formulation, found by Einstein in 1928 \cite{Einstein:1925, Einstein:1928a, Einstein:1928b, Einstein:1930, Unzicker:2005}, the gravitational degrees of freedom are carried by torsion, with curvature and non-metricity put to zero by taking the Weitzenböck connection. In the third member of the family, dubbed symmetric teleparallel theory, found by Nester and Yo in 1998 \cite{Nester:1998}, gravity is described by the non-metricity tensor, with the Riemann tensor and torsion put to zero by choosing a pure gauge connection.

The full Ricci scalar $R$ is the sum of the Ricci scalar $\mathring R$ formed from the Levi--Civita connection, the torsion scalar $T$ and the non-metricity scalar $Q$, as well as terms that mix torsion and non-metricity, and a total derivative. As curvature and non-metricity are zero in the usual teleparallel formulation, the torsion scalar is (up to a sign and a total derivative) equal to the Levi--Civita Ricci scalar. Similarly, in the symmetric teleparallel case the non-metricity scalar is equal to (minus) the Levi--Civita Ricci scalar plus a total derivative. So if the actions in the three cases are defined by using only the Ricci scalar, the torsion scalar or the non-metricity scalar, they are identical up to signs and total derivatives. The theories are therefore equivalent, at least if matter does not couple to the connection. The teleparallel and the symmetric teleparallel formulation are not modified theories of gravity, but alternatives on equal footing with the metric formulation as the starting point for extension and quantisation.

Once the actions are extended (no work has been done on quantisation in the teleparallel and symmetric teleparallel cases), the equivalence can be broken, with the three formulations branching off in different directions.
For comparison, consider the Palatini formulation (also called the metric-affine formulation), found by Einstein in 1925 \cite{Einstein:1925, ferraris1982}. There the metric and the connection are independent variables. The Palatini formulation is equivalent to the metric formulation (and to the other two formulations discussed above) for the Einstein--Hilbert action with minimally coupled matter, but becomes physically distinct when the gravitational action is more complicated \cite{Buchdahl:1960, Buchdahl:1970, ShahidSaless:1987, Flanagan:2003a, Flanagan:2003b, Sotiriou:2006, Sotiriou:2008, Olmo:2011, Borunda:2008, Querella:1999, Cotsakis:1997, Jarv:2018, Conroy:2017, Li:2007, Li:2008, Exirifard:2007, Enckell:2018b} or matter couples to the connection \cite{Lindstrom:1976a, Lindstrom:1976b, Bergh:1981, Bauer:2008, Bauer:2010, Koivisto:2005, Rasanen:2017, Enckell:2018a, Markkanen:2017, Kozak:2018}, as is the case in Higgs inflation \cite{Bezrukov:2007, Bauer:2008, Bauer:2010, Rasanen:2017, Enckell:2018a, Markkanen:2017, Rasanen:2018a, Rubio:2019}.
In the metric theory, quantum corrections induce higher order curvature terms, and new geometric terms formed from torsion and non-metricity are similarly expected to arise in the teleparallel and the symmetric teleparallel formulation. The possibilities for extension are wider than in the metric case, and modifications such as new general relativity \cite{PhysRevD.19.3524, PhysRevD.24.3312, Jimenez:2019tkx}, newer general relativity \cite{BeltranJimenez:2017}, $f(T)$ gravity \cite{Ferraro:2006jd, Ferraro:2008ey, Bengochea:2008gz, Cai:2015emx}, $f(Q)$ gravity \cite{BeltranJimenez:2017} and others \cite{Bahamonde:2017wwk, Jimenez:2019b} have been studied, and not all of the theories are yet fully understood.

\para{Higgs inflation.}

A particular case where the equivalence between different formulations of general relativity is broken is when a scalar field couples directly to the relevant gravitational scalar quantity (Ricci, torsion or non-metricity scalar). In the Standard Model of particle physics the Higgs field (and only the Higgs field) can couple to the gravitational scalars with a dimension 4 term, unsuppressed by a new mass scale. In the metric formulation, quantum corrections generate a direct coupling between the Ricci scalar and the Higgs field \cite{Callan:1970}. Such a term is the key ingredient for Higgs inflation in both the metric \cite{Bezrukov:2007} and the Palatini \cite{Bauer:2008} formulation. (For reviews of Higgs inflation, see \cite{Bezrukov:2013, Bezrukov:2015, Rubio:2018}.) Similarly, a direct coupling of the Higgs to the torsion scalar is expected in the teleparallel formulation. We can then ask whether the direct coupling can be used to distinguish between the metric and the teleparallel formulation, as it distinguishes between the metric and the Palatini formulation. Different formulations can also lead to different conclusions in the quantum theory, as happens for perturbative unitarity in the metric and the Palatini formulation \cite{Barbon:2009, Burgess:2009, Burgess:2010zq, Lerner:2009na, Lerner:2010mq, Hertzberg:2010, Bauer:2010, Bezrukov:2010, Bezrukov:2011a, Calmet:2013, Weenink:2010, Lerner:2011it, Prokopec:2012, Xianyu:2013, Prokopec:2014, Ren:2014, Escriva:2016cwl, Fumagalli:2017cdo, Gorbunov:2018llf, Ema:2019}.

Direct coupling of a scalar field to the torsion scalar \cite{Sotiriou:2010mv, Geng:2011aj, Geng:2011ka, Izumi:2013dca, Wei:2011yr, Yang:2010ji, Gu:2012ww, Geng:2012vn, Dong:2012pp, Xu:2012jf, Jamil:2012vb, Jamil:2012ck, Banijamali:2012nx, Bamba:2013jqa, Geng:2013uga, Kucukakca:2013mya, Otalora:2013tba, Li:2013oef, Otalora:2013dsa, Sadjadi:2013nb, Skugoreva:2014ena, Chen:2014qsa, Fazlpour:2014qaa, Cai:2015emx, Bahamonde:2015hza, Jarv:2015odu, Abedi:2015cya, Wright:2016ayu, Geng:2016yke, MohseniSadjadi:2016ukp, Sadjadi:2016kwj, Abedi:2017jqx, Gecim:2017hmn, Wu:2016dkt, Hohmann:2017duq, Bahamonde:2017ize, Bahamonde:2018miw, DAgostino:2018ngy, Abedi:2018lkr, Hohmann:2018rwf, Hohmann:2018vle, Akbarieh:2018oie, Hohmann:2018dqh, Hohmann:2018ijr, Golovnev:2018wbh, Bahamonde:2019gjk, Bahamonde:2019shr, Jarv:2019ctf, Gonzalez-Espinoza:2019ajd} and inflation in teleparallel gravity \cite{Ferraro:2006jd, Ferraro:2008ey, Wu:2011kh, Bamba:2012mi, Bamba:2012vg, Jamil:2013nca, Bamba:2013fta, Bamba:2013rra, Bamba:2013jqa, Bamba:2014zra, Bamba:2014eea, Nashed:2014lva, Behboodi:2014tda, Kofinas:2014daa, Hanafy:2014ica, Nashed:2014vsa, Harko:2014sja, Hanafy:2014bsa, Nashed:2015wso, Ganiou:2015aja, Nassur:2015zba, Cai:2015emx, Hanafy:2015lda, Rezazadeh:2015dza, Bamba:2016gbu, Bamba:2016wjm, Wu:2016dkt, Gecim:2016gqx, ElHanafy:2017sih, Awad:2017ign, Rezazadeh:2017edd, Akbarieh:2018oie, Goodarzi:2018feh, Abadji:2019cie, Gonzalez-Espinoza:2019ajd} have been studied in a number of papers. It has always been assumed that non-metricity vanishes, which in the tetrad formalism is known as the tetrad postulate. However, the condition that non-metricity vanishes is not invariant under changes in field space coordinates. Just as the geometric degrees of freedom can be shuffled between curvature, non-metricity and torsion, these three aspects of geometry can be partially transferred into modifications of the scalar field kinetic term and potential \cite{Kijowski:2016, Afonso:2018a, Afonso:2018c, Afonso:2017, Iosifidis:2018zjj, Iosifidis:2018zwo, Rasanen:2018b}.

We consider the teleparallel formulation with a scalar field $\varphi$ that has a direct coupling $F(\varphi)$ to the torsion scalar. We also include an arbitrary function $P(\varphi)$ in the tetrad postulate to take into account that it can be taken to hold in different frames. We further include a function $G(\varphi)$ that couples the torsion vector to the derivative of the scalar field. This coupling turns out to be crucial for the properties of the theory.  We show that if the three coupling functions $F$, $G$ and $P$ do not satisfy a specific relation, the scalar field does not generate scalar perturbations in linear theory. It thus cannot act as the inflaton. In theories where this relation is satisfied, depending on the relation between the three functions, we can get the same result as in the metric or the Palatini formulation, or a new kind of effective potential. Applying the results to Higgs inflation, we show that the tensor-to-scalar ratio can be larger than in the metric or the Palatini case. We demonstrate that our result applies to $f(T)$ type actions as they can by a change of field coordinates be transformed into an action that is linear in $T$ but has a scalar field. 

In \sec{sec:form} we go through the teleparallel formulation of gravity, in \sec{sec:scalar} we add a scalar field, derive a condition for the existence of linear scalar perturbations, and consider Higgs inflation, $f(T)$ theories and their generalisations. In \sec{sec:disc} we discuss our results and in \sec{sec:conc} we summarise our findings.

\section{Teleparallel formulation of gravity} \label{sec:form}

\subsection{Tetrads}

We consider the teleparallel formulation of general relativity using tetrads. A tetrad $e^A{}_\alpha$ is a set of four fields that forms a basis for the tangent space at each point of spacetime, providing a soldering of the spacetime manifold and the tangent space.  We denote spacetime coordinate indices with Greek letters from the beginning of the alphabet ($\alpha, \beta, \ldots$), tangent space indices with uppercase Latin letters from the beginning of the alphabet ($A, B, \ldots$), and spatial indices with lowercase Latin letters from the middle of the alphabet ($i, j, \ldots$). We take the basis to be orthonormal with respect to the metric $g_{\alpha\beta}$,
\begin{dmath}
  g_{\alpha\beta} = \eta_{A B} e^A{}_\alpha e^B{}_\beta \ ,
\end{dmath}
where $\eta_{A B}=\diag(-1,1,1,1)$ is the Minkowski metric. The inverse tetrad $e_A{}^\alpha$ is defined so that
\bea
  e_A{}^\alpha e^A{}_\beta = \delta^\alpha{}_\beta \ , \qquad e_A{}^\alpha e^B{}_\alpha = \delta^B{}_A \ .
\eea

As the metric can be written in terms of the tetrad, but not vice versa, the tetrad is more fundamental than the metric. Tetrads are needed to describe fermions in curved spacetime. The curvature-free metric-compatible connection used in the teleparallel formulation cannot be written in closed form in terms of the metric, tetrads are also needed for that. (Although teleparallel gravity can be formulated in terms of the metric and the affine connection using Lagrange multipliers, or equivalently in terms of the metric and an auxiliary transformation matrix that parametrises the curvature-free connection \cite{Kopczynski:1982, HEHL19951, Blagojevic:2000kp, Blagojevic:2002qm, Blagojevic:2002, Obukhov:2002tm, Nester:2017wau, BeltranJimenez:2018vdo, Jimenez:2019b}.) 

\subsection{Curvature, non-metricity and torsion}

An arbitrary affine connection can be written as
\bea \label{Gamma}
  \Gamma^\c_{\a\b} = \mathring\Gamma^\c_{\a\b} + L^\c{}_{\a\b} \ ,
\eea
where $\mathring\Gamma^\c_{\a\b}$ is the Levi--Civita connection of the metric $g_{\a\b}$, and $L^\c{}_{\a\b}$ is called the distortion (or the deformation) tensor. We denote quantities defined with the Levi--Civita connection with $\mathring{}$. The distortion tensor can be further decomposed as 
\begin{dmath}
  L^\c{}_{\a\b} = J^\c{}_{\a\b} + K^\c{}_{\a\b} \ ,
\end{dmath}
where $J^\c{}_{\a\b}$ is the disformation tensor and $K^\c{}_{\a\b}$ is the contortion tensor, defined as
\bea \label{L}
  J_{\a\b\c} &\equiv& \ha \left( Q_{\a\b\c}  - Q_{\c\a\b} - Q_{\b\a\c} \right) \ , \qquad K_{\a\b\c} \equiv \ha ( T_{\a\b\c} + T_{\c\a\b} + T_{\b\a\c} ) \ ,
\eea
where $Q_{\a\b\c}$ is the non-metricity and $T_{\a\b\c}$ is the torsion, defined as
\bea \label{TQ}
  Q_{\c\a\b} \equiv \nabla_\c g_{\a\b} \ , \qquad T^\c{}_{\a\b} &\equiv& 2 \Gamma^{\c}_{[\a\b]} \ .
\eea
Note that $Q_{\c\a\b}=Q_{\c(\a\b)}$, $J_{\a\b\c}=J_{\a(\b\c)}$ and $K^\c{}_\a{}^\b=K^{[\c}{}_\a{}^{\b]}$. The non-metricity vectors are defined as $Q^\alpha{} \equiv Q^{\a\b}{}_\b$ and $\hat{Q}^\alpha \equiv Q^{\b\a}{}_\b$, and the torsion vector is defined as $T^\alpha \equiv T^{\beta\alpha}{}_\beta$.

The Riemann tensor is
\bea \label{Riemann}
  R^\alpha{}_{\beta\gamma\delta} &\equiv& \Gamma^\alpha_{\delta\beta,\gamma} - \Gamma^\alpha_{\gamma\beta,\delta} + \Gamma^\alpha_{\gamma\mu} \Gamma^\mu_{\delta\beta} - \Gamma^\alpha_{\delta\mu} \Gamma^\mu_{\gamma\beta} \el
  &=& \mathring R^{\a}{}_{\b\c\d} + 2 \mathring \nabla_{[\c} L^\a{}_{\d]\b} + 2 L^\a{}_{[\c|\mu|} L^\mu{}_{\d]\b} \ ,
\eea
where on the second line we have decomposed the Riemann tensor into the Levi--Civita and distortion contributions. The Ricci scalar can be written as
\bea \label{R}
  R &\equiv& R^{\a\b}{}_{\a\b} = \mathring R + Q + T + \mathring\nabla_\a ( Q^\a - \hat Q^\a + 2 T^\a ) - ( Q_\a - \hat Q_\a ) T^\a + Q_{\a\b\c} T^{\c\a\b} \ ,
\eea
where we have used \re{Gamma}--\re{Riemann} to separate the contributions of curvature, non-metricity and torsion. The non-metricity and torsion scalars are defined as
\bea \label{QT}
  Q &\equiv& \frac{1}{4} Q_{\a\b\c} Q^{\a\b\c} - \ha Q_{\a\b\c} Q^{\c\a\b} - \frac{1}{4} Q_\a Q^\a + \ha Q_\a \hat Q^\a \el
  T &\equiv& \frac{1}{4} T_{\a\b\c} T^{\a\b\c} - \ha T_{\a\b\c} T^{\c\a\b} - T_\a T^\a \ ,
\eea
respectively. In teleparallel gravity, the curvature and non-metricity are usually taken to be zero. In that case we have $T=-\mathring R$, neglecting total derivatives. Similarly in the symmetric teleparallel case, torsion is zero and $Q=-\mathring R$. Actions linear in $R$, $Q$ or $T$ are equivalent: the gravitational physics can be shifted between $R$, $-Q$ and $-T$. However, the equivalence is broken when these geometrical scalars are directly coupled to a scalar field (so the total derivative terms contribute) or when the action is non-linear in the appropriate scalar (as is well known from $f(T)$ and $f(Q)$ theories) \cite{Bahamonde:2015zma, ElHanafy:2017sih, Krssak:2018ywd}. We will consider both scalar field coupling and non-linearity, and take into account that non-metricity and torsion can be simultaneously non-zero, in which case the cross-terms in \re{R} contribute.

\subsection{Affine connection and spin connection}

In the metric formulation, the assumption that non-metricity and torsion vanish determines the connection uniquely to be the Levi--Civita connection. In the teleparallel formulation, gravity is carried by torsion, and curvature and non-metricity vanish. The unique connection for which the Riemann tensor and the non-metricity tensor vanish is the Weitzenböck connection
\bea \label{WB}
  \Gamma^\c_{\a\b} = e_A{}^\c \pat_\a e^A{}_\b + e_A{}^\c \omega^A{}_{\a B} e^B{}_\b \ ,
\eea
where $\omega^A{}_{\a B}$ is the spin connection, which defines the covariant derivative $\mathcal{D}_\b$ in the tangent space as $\mathcal{D}_\b e^A{}_\a\equiv\pat_\b e^A{}_\a+\omega^A{}_{\b B} e^B{}_\a$.

The Weitzenböck connection fixes the spacetime affine connection, but we also have to account for the structure of the tangent space. Like the spacetime, the tangent space is flat, and it has zero non-metricity and torsion, so it is Minkowski spacetime. It is therefore possible to choose the tangent space coordinates (called a frame\footnote{The term {\em frame} has two different meanings. It refers both to the choice of coordinates in tangent space and the choice of coordinates in the space of fields formed by the metric, connection and matter fields. Which is meant should be clear from the context.}) so that the spin connection vanishes, though other choices may be more convenient. All tangent space frames are related by local Lorentz transformations $\Lambda^A{}_B(x)$ that leave the tangent space metric invariant,
\bea \label{Lorentz}
  \eta_{AB} \to \Lambda^C{}_A(x) \Lambda^D{}_B(x) \eta_{CD} = \eta_{AB} \ ,
\eea
and transform the tetrad and the spin connection as
\bea
  e^A{}_\a &\to& \Lambda^A{}_B e^B{}_\a \\
  \label{spint} \omega^{A}{}_{\alpha B} &\to& \Lambda^{A}{}_{C} \omega^C{}_{\alpha D} (\Lambda^{-1})^{D}{}_{B} - \Lambda^{A}{}_C{}_{,\alpha} (\Lambda^{-1})^C{}_{B} \ .
\eea
Using \re{spint} and the fact that there is a frame where the spin connection vanishes, the spin connection in a general frame can be written in terms of the local Lorentz transformation as
\begin{dmath} \label{spinl}
  \omega^{A}{}_{\alpha B} = - \Lambda^{A}{}_C{}_{,\alpha} (\Lambda^{-1})^C{}_{B} \ .
\end{dmath}
The spin connection contains 6 degrees of freedom, all of which are pure gauge. The tetrad has 16 components: 10 describe the spacetime metric, and 6 correspond to the choice of frame in tangent space. The information about the choice of frame is also carried by the spin connection, so there are 6 redundant gauge degrees of freedom: we can specify the frame using either the tetrad or the spin connection \cite{Krssak:2015oua, Hohmann:2018rwf, Krssak:2018ywd, Hohmann:2019nat, Jarv:2019ctf}.

\section{Teleparallel gravity coupled to a scalar field} \label{sec:scalar}

\subsection{The action}

We consider teleparallel gravity coupled non-minimally to a scalar field and minimally to other matter, with the action
\begin{dmath}[label={Taction}]
  S = \int \dd[4]{x} e \left[ - \ha F(\varphi) T - G(\varphi) \mathring\nabla_\alpha T^\alpha - \ha K(\varphi) g^{\a\b} \pat_\a \varphi \pat_\b \varphi - V(\varphi) \right] + S_{\textrm{(m)}}(\Psi, \varphi, e^A{}_\a, \mathring\omega^A{}_{\a B}) \ ,
\end{dmath}
where $e\equiv\det(e^A{}_\a)$, and we have included an arbitrary kinetic function $K$ and potential $V$. The matter part of the action, $S_{\textrm{(m)}}$, can depend on $\varphi$, other matter degrees of freedom collectively denoted by $\Psi$, the tetrad $e^A{}_\a$ and the Levi--Civita spin connection $\mathring\omega^A{}_{\a B}$ (needed for fermions and defined with \re{WB} using the Levi--Civita affine connection), but it does not depend on the affine connection $\Gamma^\gamma_{\alpha\beta}$ nor on the spin connection $\omega^A{}_{\alpha B}$. As we ignore boundary terms, the torsion vector coupling function $G$ is only defined up to an arbitrary additive constant. The minimally coupled teleparallel case is obtained for $F=1, G=0$ (we choose units such that Planck mass is unity).

In teleparallel gravity, it is usually assumed that the tetrad postulate holds, \ie the tetrad is covariantly constant. Together with vanishing curvature, this gives the Weitzenböck connection. However, the tetrad postulate is not invariant under changes of field space coordinates, in particular under conformal transformations. We thus include a coupling function $P(\varphi)$ in the tetrad postulate: we assume that (in the Jordan frame in which \re{Taction} is written) we have
\begin{dmath} \label{tetradpos}
  \nabla_\b [ P(\varphi) e^A{}_\a ] = 0 \ ,
\end{dmath}
where $\nabla$ is the total covariant derivative that acts both on spacetime and tangent space indices, $\nabla_\b e^A{}_\a = \pat_\b e^A{}_\a+\omega^A{}_{\b B} e^B{}_\a-\Gamma^\c_{\b\a} e^A{}_\c$. Note that $P$ is defined only up to an arbitrary multiplicative non-zero constant. As the curvature is zero, the connection is Weitzenböck for the tetrad $P e^A{}_\a$. In terms of the tetrad $e^A{}_\a$ the connection is, according to \re{WB},
\bea \label{WBP}
  \Gamma^\c_{\a\b} = e_A{}^\c \pat_\a e^A{}_\b + e_A{}^\c \omega^A{}_{\a B} e^B{}_\b + \delta^\c{}_\b \pat_\a\ln P\ .
\eea
In terms of the metric, \re{tetradpos} corresponds to $\nabla_\c ( P^2 g_{\a\b} )=0$, \ie
\bea \label{Q}
  Q_{\c\a\b} = - 2 g_{\a\b}\pat_\c\ln P = - 2 \frac{P'}{P} g_{\a\b} \pat_\c \varphi \ ,
\eea
where prime denotes derivative with respect to $\varphi$.

Using the decomposition \re{R}  of the Ricci scalar in the action \re{Taction}, inputting $R=0$ and the non-metricity tensor \re{Q}, and dropping a boundary term, we obtain the action in terms of the Levi--Civita curvature scalar,
\begin{dmath}[label={LCaction}]
  S = \int \dd[4]{x} e \left[ \ha F \mathring R  + \left( 2 F \frac{P'}{P} + G' - F' \right) T^\a \pat_\a \varphi - \ha \left\{ K + 6 F \frac{P'}{P} \left( \frac{P'}{P} - \frac{F'}{F} \right) \right\} g^{\a\b} \pat_\a \varphi \pat_\b \varphi - V \right] + S_{\textrm{(m)}}(\Psi, \varphi, e^A{}_\a, \mathring\omega^A{}_{\a B}) \ .
\end{dmath}
We get two qualitatively different theories, depending on whether or not the derivative coupling to the torsion vector on the first line vanishes.

\subsection{Zero torsion vector coupling} \label{sec:zero}

\subsubsection{Einstein frame action}

Let us first consider the case when the derivative coupling to the torsion vector in the action \re{LCaction} vanishes,
\bea \label{cons}
  \frac{P'}{P} = \frac{F'-G'}{2 F} \ .
\eea
The action \re{LCaction} then reads
\begin{dmath}[label={novectoraction}]
  S = \int \dd[4]{x} e \left[ \ha F \mathring R - \ha \left( K - \frac{3}{2} \frac{ F'{}^2 - G'{}^2 }{F} \right) g^{\a\b} \pat_\a \varphi \pat_\b \varphi - V \right] + S_{\textrm{(m)}}(\Psi, \varphi, e^A{}_\a, \mathring\omega^A{}_{\a B}) \ .
\end{dmath}
Doing the conformal transformation $e^A{}_\a\to F^{-1/2} e^A{}_\a$, taking into account $e\to F^{-2} e$ and $\mathring R\to F \left( \mathring R - \frac{3}{2} g^{\a\b} \pat_\a \ln F\pat_\b \ln F + 3 \mathring\Box\ln F \right)$, and dropping a boundary term, we get
\begin{dmath}[label={minaction}]
  S = \int \dd[4]{x} e \left[ \ha \mathring R - \ha \left( \frac{K}{F} + \frac{3}{2} \frac{ G'{}^2 }{F^2} \right) g^{\a\b} \pat_\a \varphi \pat_\b \varphi - \frac{V}{F^2} \right] + S_{\textrm{(m)}}[\Psi, \varphi, F^{-1/2} e^A{}_\a, \mathring\omega^A{}_{\a B}(F)] \ ,
\end{dmath}
where $\mathring\omega^A{}_{\a B}(F)$ is the Levi--Civita spin connection corresponding to the tetrad $F^{-1/2} e^A{}_\a$. The action has been reduced to the usual Einstein--Hilbert term of general relativity written with the Levi--Civita connection plus a minimally coupled scalar field. The effects of the non-minimal couplings and the tetrad postulate coupling have been transferred to the scalar field kinetic term and potential.

If the tetrad postulate holds in the original Jordan frame, $P'=0$, we have $G=F$, and the action is identical to that of a non-minimally coupled scalar field in the metric formalism \cite{Futamase:1989, Salopek:1989, Fakir:1990, Makino:1991, Fakir:1992, Kaiser:1994, Komatsu:1999, Bezrukov:2007}. We get the same action for $G=-F$, which corresponds to $P=F$. This case is halfway between the metric and the Palatini formalism, and non-metricity is non-zero in both in the original Jordan frame and in the Einstein frame. The case when the tetrad postulate holds in the Einstein frame, $P^2=F$, corresponds to $G=0$, \ie the absence of a coupling to $\mathring\nabla_\a T^\a$ in the original action \re{Taction}. The final action \re{minaction} is then identical to that of a non-minimally coupled scalar field in the Palatini formalism \cite{Bauer:2008}.

Finally, if $|G'|\neq F'\neq0$, we obtain a new kind of a modification to the kinetic term. For example, if $F=1$ and $G'{}^2\gg K$, the field with a canonical kinetic term is \mbox{$\chi=\pm\sqrt{\frac{3}{2}}G(\varphi)+\chi_0$}, so the potential becomes $V[G^{-1}(\pm\sqrt{\frac{2}{3}}\chi\mp\sqrt{\frac{2}{3}}\chi_0)]$, where $G^{-1}$ is the inverse function of $G$. By choosing $G(\varphi)$ appropriately, it is then possible to obtain inflation almost regardless of the shape of the potential $V(\varphi)$. If $1/G'(\varphi)\to0$ suitably fast as $\varphi\to\varphi_0$ for some finite $\varphi_0$, the kinetic term makes any smooth potential $V(\varphi)$ suitably flat for inflation when written in terms of $\chi$. This is the $\a$-attractor mechanism \cite{Ferrara:2013, Kallosh:2013, Galante:2014}.

\subsubsection{The Higgs case} \label{sec:Higgs}

Let us now take $\varphi$ to be the Standard Model Higgs, in which case it appears only in even powers and the tree-level potential is
\bea \label{V}
  V = \frac{\lambda}{4} ( \varphi^2 - v^2 )^2 \ .
\eea
If we consider only terms up to dimension 4 not just in the potential but also in the coupling functions, we have (recalling that $G$ is defined only up to an additive constant)
\bea
  K &=& K_0 \el
  F &=& F_0 ( 1 + \xi \varphi^2 ) \el
  G &=& G_1 \varphi^2 \ ,
\eea
where $K_0$, $F_0$ and $G_1$ are constants. Note that unless $G'=(1-n)F'$, where $n$ is a positive integer ($n=0$  and $n=1$ correspond to the metric and the Palatini case, respectively), the function $P^2$ given by \re{cons} is not polynomial.

Let us first consider the usual inflationary regime $\xi\varphi^2\gg1$, where the effective potential $U=V/F^2$ is asymptotically flat. The transformation between $\varphi$ and the canonical field $\chi$ is (neglecting constant rescalings of $\lambda$ and $v$ by putting $K_0=F_0=1$)
\bea
  \frac{\rmd\chi}{\rmd\varphi} = \pm \sqrt{ \frac{1 + ( \xi +  6 G_1^2 ) \varphi^2 }{ ( 1 + \xi \varphi^2 )^2 } } \simeq  \sqrt{ \frac{ \xi +  6 G_1^2 }{\xi^2} } \frac{1}{\varphi} \ ,
\eea
so the potential is exponentially flat in terms of $\chi$. To leading order, the spectral index $n_s=1-2/N$ depends only on the number of e-folds $N$, and is in good agreement with observations. The amplitude of scalar perturbations $A_s$ and the tensor-to-scalar ratio $r$ depend on $\xi$ and $G_1$ (see \eg \cite{Rasanen:2017}),
\bea \label{Higgsr}
  A_s &=& \frac{N^2}{12 \pi^2} \frac{ \lambda }{ \xi + 6 G_1^2 } \el
  r &=& \frac{12}{N^2} \frac{\xi+6 G_1^2}{6 \xi^2} = \frac{\lambda}{6\pi^2 A_s \xi^2} = 8\times10^6 \frac{\lambda}{\xi^2} \ ,
\eea
where in the last equality we have input the observed value $A_s=2\times10^{-9}$ \cite{Planck2018}. If $\xi\gg 6 G_1^2$, the situation reduces to the Palatini case with $\lambda/\xi=10^{-10}$ and $r=\frac{2}{\xi N^2}\ll1$ \cite{Bauer:2008}. In the opposite limit $6 G_1^2\gg\xi$, we get the new result $\lambda/G_1^2=6\times10^{-10}$ and $r=\frac{12}{N^2} \frac{G_1^2}{\xi^2}$. Two things are noteworthy. First, the value of $\xi$ is decoupled from the amplitude of the perturbations: the small value of $A_s$ does not imply a large value of $\xi$ (but a large value of $G_1$). Second, the tensor-to-scalar ratio can be enhanced compared to the Palatini case as much as desired by adjusting $G_1^2/\xi^2$.

In the case $\xi\varphi^2\ll1$ (again taking $K_0=F_0=1$), the shape of the potential is only modified by $G$. In the limit $G_1\varphi^2\ll1$, we have the unmodified $\varphi^4$ Higgs potential. In the limit $G_1\varphi^2\gg1$, the potential in terms of the canonical field becomes 
\bea
  U = \frac{\lambda}{6 G_1^2} \chi^2 \ ,
\eea
and in between the potential interpolates between the quadratic and the quartic case. The quadratic potential gives a good fit to observations, except for the predicted tensor-to-scalar ratio $r\approx0.1$, which is too high \cite{Planck2018}. In the Palatini formulation, $r$ can be brought down by including a $R^2$ term in the action \cite{Enckell:2018b}. However, in \sec{sec:ft} we will see that a similar mechanism may not work in the teleparallel case, because if the action is non-linear in $T$, the derivative coupling to the torsion vector will not vanish unless $G'=0$. Let us see what happens when the coupling to the torsion vector is not zero.

\subsection{Non-zero torsion vector coupling} \label{sec:nonzero}

\subsubsection{Tetrad equation of motion}

We now consider the case when the derivative coupling to the torsion vector in the action \re{LCaction} does not vanish. Variation of the action \re{Taction} with respect to $e^A{}_\a$, taking into account the tetrad postulate \re{tetradpos}, gives the equation of motion
\begin{dmath}[label={eom}]
- \ha g_{\alpha\beta} F T + F S^{\gamma}{}_\beta{}^{\delta} T_{\gamma\alpha\delta} + g_{\alpha\beta} G_{,\gamma} T^\gamma - G_{,\alpha} T_\beta - G_{,\beta} T_\alpha - G_{,\gamma} T_\beta{}^{\gamma}{}_{\alpha} - e^{-1} g_{\gamma\beta} e^A{}_\alpha \left( e F S_\delta{}^{\gamma\mu} e_A{}^\delta + e G_{,\delta} g^{\delta\mu} e_A{}^\gamma - e G_{,\delta} g^{\delta\gamma} e_A{}^\mu \right){}_{,\mu} + \left( F S_{\gamma\beta}{}^{\delta} e^B{}_\alpha e_A{}^\gamma + G_{,\gamma} g^{\gamma\delta} g_{\mu\beta} e^B{}_\alpha e_A{}^\mu - G_{,\beta} e^B{}_\alpha e_A{}^\delta \right) \omega^A{}_{\delta B} + \left( F S_{\alpha\beta}{}^{\gamma} + G_{,\delta} g^{\delta\gamma} g_{\alpha\beta} - G_{,\beta} \delta^\gamma{}_\alpha \right) \frac{P_{,\gamma}}{P} = \ha g_{\alpha\beta} g^{\gamma\delta} K \varphi_{,\gamma} \varphi_{,\delta} - K \varphi_{,\alpha} \varphi_{,\beta} + g_{\alpha\beta} V + T_{\alpha\beta}^{\textrm{(m)}} \ ,
\end{dmath}
where $S_{\a\b\c}\equiv K_{\b\a\c} + g_{\a\b} T_\c - g_{\a\c} T_\b=S_{\a[\b\c]}$ is the superpotential, and \sloppy $T_{\a\b}^{\textrm{(m)}} \equiv -e^{-1} e^A{}_{\alpha} g_{\beta\gamma} \fdv{S_{(\text{m})}}{e^A{}_\gamma} = T_{(\a\b)}^{\textrm{(m)}}$ is the energy-momentum tensor of the matter fields in $S_{\textrm{(m)}}$.

The spin connection (properly constrained to depend on the Lorentz transformation according to \re{spinl}) can also be varied. However, as the action is invariant under local Lorentz transformations, which affect both the tetrad and the spin connection, any variation of the spin connection can be transformed into a variation of the tetrad, so the equation of motion of the spin connection is contained in the tetrad equation of motion \cite{Golovnev:2017dox, Hohmann:2017duq, Hohmann:2018vle, Hohmann:2018dqh, Hohmann:2018rwf}. Hence the spin connection equation of motion is given by the antisymmetric part of \re{eom}. Because of this redundancy, it would be consistent to set the spin connection to zero in the action by fixing the local Lorentz transformations. This would not lose any equations of motion or change the number of physical degrees of freedom \cite{Blixt:2019mkt} (for the case of redundant degrees of freedom in the metric, see \cite{Lagos:2013}). We leave the frame free to make it transparent that the results do not depend on the choice of frame.

\subsubsection{Perturbed FRW case}

We consider a linearly perturbed Friedmann--Robertson--Walker (FRW) spacetime with a spatially flat background, relevant for inflationary and post-inflationary universes. The background tetrad can be chosen to be diagonal, so the full tetrad reads, writing down the most general perturbation possible,
\begin{dmath}[label={tetradpert}]
  e^A{}_\alpha = \bar e^A{}_\a + \d e^A{}_\a = a \begin{bmatrix*}[l] 1 + A & \quad - \beta_{,i} + \beta_i \\ - \zeta_{,i} + \zeta_i & \quad (1 - \psi) \delta_{i j} - \epsilon_{ij}{}^k \left( \sigma_{,k} - \sigma_k \right) + E_{,i j} - E_{(i,j)} + E_{i j} \end{bmatrix*}
\end{dmath},
where overbar denotes background quantities and $\d$ denotes perturbations, $a$ is the cosmological scale factor, $A,  E, \psi, \beta$ and $\zeta$ are scalars, $\sigma$ is a pseudoscalar, $E_i, \beta_i$ and $\zeta_i$ are transverse vectors, $\sigma_i$ is a transverse pseudovector and $E_{ij}$ is a transverse traceless tensor. We are interested in scalar perturbations, which at the linear level mix only with other scalars and pseudoscalars. Furthermore, around the FRW background, pseudoscalars do not mix with the scalars at linear order, as there are no background spatial pseudovectors. We therefore drop the pseudoscalar, vectors, the pseudovector and the tensor. The metric corresponding to \re{tetradpert} is then
\bea \label{metricpert}
  \rmd s^2 = a(\eta)^2 \left\{ - ( 1+ 2 A ) \rmd\eta^2 - 2 ( \zeta - \beta ),_{i} \rmd\eta\rmd x^i + [ (1 - 2 \psi) \delta_{i j} + 2 E_{,i j} ] \rmd x^i \rmd x^j \right\} \ ,
\eea
where $\eta$ is conformal time. We see that $\zeta-\beta$ corresponds to metric perturbations, so $\zeta+\beta$ is a gauge degree of freedom that corresponds to perturbations in the local Lorentz transformations.

It is consistent with the choice of background tetrad \re{tetradpert} to set the background spin connection to zero, $\bar\omega^A{}_{\a B}=0$ \cite{Hohmann:2019nat}. As the perturbations of the tetrad \re{tetradpert} are the most general possible, we could also set the perturbed spin connection to zero. We keep the perturbation of the spin connection general and write it in terms of the perturbed Lorentz transformation
\begin{dmath}
  \Lambda^A{}_B = \delta^A{}_B + \d \Lambda^A{}_B \ ,
\end{dmath}
where we have taken into account that the background spin connection is zero, so $\bar\Lambda^A{}_B$ is constant. From the condition \re{Lorentz} (which says that $\Lambda^A{}_B$ is a local Lorentz transformation) it follows that $\d\Lambda_{AB}=-\d\Lambda_{BA}$. The six independent components correspond to three local boosts $b_i$ and three local rotations $r_i$. We write $\d\Lambda_{AB}$ as
\begin{dmath}
  \d\Lambda_{AB} = \begin{bmatrix*}[l] 0 & \quad b_i \\ - b_i & \quad \epsilon_{ij}{}^k r_k \end{bmatrix*} \ ,
\end{dmath}
so the full local Lorentz transformation is
\begin{dmath} \label{Lambdapert}
  \Lambda^A{}_B = \begin{bmatrix*}[l] 1 & \quad - b_i \\ -b_i & \quad \delta_{ij} + \epsilon_{ij}{}^k r_k \end{bmatrix*} \ ,
\end{dmath}
and according to \re{spinl} the spin connection is
\begin{dmath}
  \omega^A{}_{\alpha B} = \begin{bmatrix*}[l] 0 & \quad b_{i,\alpha} \\ b_{i,\alpha} & \quad - \epsilon_{ij}{}^k r_{k,\alpha} \end{bmatrix*}
\end{dmath}.
In the scalar sector we consider, the perturbation of the spin connection (like the perturbation of the tetrad) contains one scalar $b$ given by $b_i= b,_i$ and one pseudoscalar $r$ given by $r_i= r,_i$.

Finally, the general energy-momentum tensor of matter other than $\varphi$ can, without loss of generality, be decomposed with respect to a four-velocity $u^\a$ as \cite{Ellis:1971}
\bea
  T^{\textrm{(m)}}_{\a\b} = ( \rho + p ) u_\a u_\b +  p g_{\a\b} + 2 u_{(\a} q_{\b)}  + \Pi_{\a\b} \ ,
\eea
where $\rho$, $p$, $q_\a$ and $\Pi_{\a\b}$ are the energy density, pressure, energy flux and anisotropic stress, respectively. The quantities $q_\a$ and $\Pi_{\a\b}$ are orthogonal to $u^\a$, and $\Pi_{\a\b}$ is traceless. As we only consider the scalar sector and linear perturbations, we have
\begin{dmath}
  T^{\textrm{(m)}\a}{}_\beta = \bar{T}^{\textrm{(m)}\a}{}_\beta + \d T^{\textrm{(m)}\a}{}_\beta
= \begin{bmatrix*}[l] - \bar{\rho} - \delta{\rho} & \quad - ( \bar{\rho} + \bar{p}) (v + q + \zeta - \beta )_{,i} \\ (\bar{\rho} + \bar{p}) (v+q)_{,i} & \quad \bar{p} \delta_{i j} + \delta{p} \delta_{i j} + \bar{p} (\Pi_{,i j} - \tfrac{1}{3} \delta_{i j} \laplacian \Pi) \end{bmatrix*} \ ,
\end{dmath}
where we have written $u^i=-a^{-1}v,_i$ and $q^i=-a^{-1} (\bar{\rho} + \bar{p})q,_i$.

Let us start with the antisymmetric part of the tetrad equation of motion \re{eom}. At the background level, it is satisfied identically due to the FRW symmetry. At first order in perturbations, it reads, given \re{Lambdapert},
\begin{dgroup*}
\begin{dmath} \label{eomas1}
\left(2 \bar{F} \frac{\dot{\bar{P}}}{\bar{P}} + \dot{\bar{G}} - \dot{\bar{F}} \right) \left[ \left( \psi + \frac{\mathcal{H}}{\dot{\bar{\varphi}}} \delta{\varphi} \right)_{,i} - \ha  \epsilon_i{}^{j k} r_{j,k} \right] = 0
\end{dmath}
\begin{dmath} \label{eomas2}
\left(2 \bar{F} \frac{\dot{\bar{P}}}{\bar{P}} + \dot{\bar{G}} - \dot{\bar{F}} \right) \epsilon_i{}^{j k} b_{j,k} = 0 \ ,
\end{dmath}
\end{dgroup*}
where dot denotes derivative with respect to $\eta$, and $\mathcal{H}=\adot/a$ is the conformal Hubble parameter. The background prefactor in \re{eomas1} and \re{eomas2} is precisely the combination that appears in the derivative coupling to the torsion vector in the action \re{LCaction}. If it vanishes, we are back to the case discussed in \re{sec:zero}, and the antisymmetric part of the equation of motion is satisfied identically. In the present case, \re{eomas2} simply gives $b_i=b,_i$ as anticipated above. Contracting \re{eomas1} with $\delta^{il}\pat_l$ leads to
\begin{dmath}
  \laplacian \left( \psi + \frac{\mathcal{H}}{\dot{\bar{\varphi}}} \delta{\varphi} \right) = 0 \ ,
\end{dmath}
which gives
\bea \label{psi}
 \psi + \frac{\mathcal{H}}{\dot{\bar{\varphi}}} \delta{\varphi} = 0 \ .
\eea
Inserting this back into \re{eomas1} gives $r_i=r,_i$. If there is no matter other than $\varphi$, the combination \re{psi} is the comoving curvature perturbation, which is identically zero.

The symmetric part of the tetrad equations of motion for the background then reads
\begin{dgroup*}
\begin{dmath}
- 3 \dot{\bar{G}} \frac{\dot{\bar{P}}}{\bar{P}} - 3 \dot{\bar{G}} \mathcal{H} - 3 \bar{F} \frac{{\dot{\bar{P}}}^{2}}{\bar{P}^2} - 6 \bar{F} \frac{\dot{\bar{P}}}{\bar{P}} \mathcal{H} - 3 \bar{F} \mathcal{H}^{2} = a^{2} \bar{\rho} - \ha \bar{K} \dot{\bar{\varphi}}^2 - a^{2} \bar{V}
\end{dmath},
\begin{dmath}
- 2 \bar{F} \frac{\dot{\bar{P}}}{\bar{P}} \mathcal{H} + 2 \dot{\bar{F}} \frac{\dot{\bar{P}}}{\bar{P}} + 2 \bar{F} \frac{\ddot{\bar{P}}}{\bar{P}} + \ddot{\bar{G}} + 2 \bar{F} \dot{\mathcal{H}} + 2 \dot{\bar{F}} \mathcal{H} - 5 \bar{F} \frac{\dot{\bar{P}}^2}{\bar{P}^2} - 3 \dot{\bar{G}} \frac{\dot{\bar{P}}}{\bar{P}} - \dot{\bar{G}} \mathcal{H} + \bar{F} \mathcal{H}^{2} = a^{2} \bar{p} - \ha \bar{K} \dot{\bar{\varphi}}^2 + a^{2} \bar{V} 
\end{dmath}.
\end{dgroup*}
We adopt the gauge $\d\varphi=E=0$, in which case \re{psi} reduces to $\psi=0$, and the perturbation of the symmetric part of the equation of motion reads
\begin{align}
&- \left( 2 \bar{F} \frac{\dot{\bar{P}}}{\bar{P}} + \dot{\bar{G}} + 2 \bar{F} \mathcal{H} \right) \laplacian \left( \zeta - \beta \right) = a^{2} \delta{\rho} + 2 a^{2} \left( \bar{\rho} - \bar{V} \right) A \label{sc-pert-1} \\
&- \left( 2 \bar{F} \frac{\dot{\bar{P}}}{\bar{P}} + \dot{\bar{G}} + 2 \bar{F} \mathcal{H} \right) A = a^{2} \left( \bar{\rho} + \bar{p} \right) \left( v + q + \zeta - \beta \right) \label{sc-pert-2} \\
&- \left( 2 \bar{F} \frac{\dot{\bar{P}}}{\bar{P}} + \dot{\bar{G}} + 2 \bar{F} \mathcal{H} \right) \dot{A} = a^{2} \delta{p} + 2 a^{2} \left( \bar{p} + \bar{V} \right) A + \frac{2}{3} a^{2} \bar{p} \laplacian \Pi \label{sc-pert-3} \\
&\bar{F} A - ( \dot{\bar{F}} + 2 \bar{F} \mathcal{H} ) ( \zeta - \beta ) - \bar{F} ( \dot\zeta - \dot\beta ) - \left( 2 \bar{F} \frac{\dot{\bar{P}}}{\bar{P}} + \dot{\bar{G}} - \dot{\bar{F}} \right) ( b - \beta ) = a^{2} \bar{p} \Pi \label{sc-pert-4} \ .
\end{align}
If there is no matter other than $\varphi$, then \eqref{sc-pert-2} yields $A = 0$ (it follows from the background equations that the left-hand side prefactor is non-zero). Then \eqref{sc-pert-1} gives $\zeta = \beta$ and \eqref{sc-pert-4} gives $\beta=b$, leaving only a gauge degree of freedom. So if there is no matter apart from $\varphi$, there are no scalar metric perturbations. This result has previously been discussed in \cite{Wu:2016dkt, Golovnev:2018wbh, Gonzalez-Espinoza:2019ajd}.

If other matter is included, the combination \re{psi} remains zero regardless of the form of the matter action $S_{\textrm{(m)}}$. The reason is that as long as the matter fields do not couple to the spin connection, they do not contribute to the spin connection equation of motion, which is equivalent to the antisymmetric part of the tetrad equation of motion. (The fact that the energy-momentum tensor is symmetric is an expression of this.) However, the comoving curvature perturbation will be sourced by other matter components, and will no longer be equal to \re{psi}, and the other metric perturbations are in general non-zero, as \eqref{sc-pert-1}--\eqref{sc-pert-4} show.
 
In summary, a single scalar field does not source linear scalar perturbations, and so does not give a working inflationary model, unless the coupling functions of the torsion scalar, torsion vector and the tetrad postulate in the original action \re{Taction} are related by the condition \re{cons} so that the derivative coupling to the torsion vector vanishes in the Einstein frame action \re{LCaction}. For example, a theory with a direct coupling only to the torsion scalar (and no non-metricity) is not viable, nor is a theory with only a derivative coupling to the torsion vector (and no non-metricity).

\subsection{Non-linearity in the torsion scalar} \label{sec:ft}

\subsubsection{The general case $f(T+2\mu\mathring\nabla_\alpha T^\alpha, \varphi)$}

One much discussed extension of the simplest action for teleparallel gravity is the theory where the gravitational action is non-linear in $T$. We first consider the general case \mbox{$f(T+2\mu\mathring\nabla_\a T^\a, \varphi)$} \cite{Bamba:2013jqa, Otalora:2014aoa, Fazlpour:2014qla, Bahamonde:2015hza, doi:10.1142/S0218271817501036, Abedi:2017jqx, FAZLPOUR20181322, Hohmann:2018rwf, Hohmann:2018vle, Hohmann:2018dqh, Hohmann:2018ijr}, where $\mu$ is a constant and $\varphi$ is a scalar field. Let us see how the action is transformed into an action that is linear in $T+2\mu\mathring\nabla_\a T^\a$ and how our results apply. We start with the action
\begin{dmath}[label={ftbaction}]
  S = \int \dd[4]{x} e \left[ \ha f(T+2\mu \mathring\nabla_\alpha T^\alpha, \varphi) - G(\varphi) \mathring\nabla_\alpha T^\alpha - \ha K(\varphi) g^{\a\b} \pat_\a \varphi \pat_\b \varphi \right] + S_{\textrm{(m)}}(\Psi, \varphi, e^A{}_\a, \mathring\omega^A{}_{\a B}) \ ,
\end{dmath}
where, as before, $\Psi$ denotes matter degrees of freedom other than $\varphi$. Note that the term $f(T+2\mu \mathring\nabla_\alpha T^\alpha, \varphi)$ includes the case where the linear terms $T$ and $\mathring\nabla_\alpha T^\alpha$ couple to the same function of $\varphi$, so now (for $\mu\neq0$) $G$ parametrises the difference between the scalar field coupling of these two linear terms. By introducing the auxiliary field $\theta$, we can write the action as \cite{Yang:2010ji, Wright:2016ayu}
\begin{dmath}[label={ftbaction-scalarfield}]
  S = \int \dd[4]{x} e \left[ - \ha F(\theta,\varphi) T - [ G(\varphi) + \mu F(\theta,\varphi) ] \mathring\nabla_\alpha T^\alpha + \ha \theta F(\theta,\varphi) + \ha f(\theta,\varphi) - \ha K(\varphi) g^{\a\b} \pat_\a \varphi \pat_\b \varphi \right] + S_{\textrm{(m)}}(\Psi, \varphi, e^A{}_\a, \mathring\omega^A{}_{\a B}) \ ,
\end{dmath}
where $F(\theta,\varphi) \equiv - \frac{\pat f}{\pat\theta}$. If $\frac{\pat^2 f}{\pat\theta^2}=0$, we have $f(\theta,\varphi)=-F(\varphi) \theta - 2 V(\varphi)$, and the action reduces to the linear case we have already considered. If $\frac{\pat^2 f}{\pat\theta^2}\neq0$, it is straightforward to verify that by varying \re{ftbaction-scalarfield} with respect to $\theta$ gives $\theta=T+2\mu\mathring\nabla_\a T^\a$ and we recover \re{ftbaction} by substituting it back into the action. Repeating now the steps of using \re{R} to write the torsion scalar in terms of the Levi--Civita Ricci scalar, non-metricity tensor and torsion vector, writing the non-metricity tensor in terms of the coupling function $P(\theta,\varphi)$ with the tetrad postulate \re{tetradpos}, and making the conformal transformation $e^A{}_\a\to F^{-1/2} e^A{}_\a$, we get the Einstein frame action
\begin{dmath}[label={ftminaction1}]
  S = \int \dd[4]{x} e \left[ \ha \mathring R -  F^{-2} T^\a \left\{ F \pat_\a \ln P^2 + \pat_\a [ G + ( \mu - 1 ) F ] \right\} - \frac{3}{4} g^{\a\b} \pat_\a \ln\frac{P^2}{F} \pat_\b \ln\frac{P^2}{F} - \ha \frac{K}{F} g^{\a\b} \pat_\a \varphi \pat_\b \varphi - U \right] + S_{\textrm{(m)}}[\Psi, \varphi, F^{-1/2} e^A{}_\a, \mathring\omega^A{}_{\a B}(F)] \ ,
\end{dmath}
where $U(\theta,\varphi)\equiv-[ \ha \theta F(\theta,\varphi) + \ha f(\theta,\varphi) ]/F(\theta,\varphi)^2$. We will not embark on a full analysis of the conditions for this theory to support scalar perturbations, and only consider some special cases.

Because $G$ is a function of $\varphi$ only, whereas $F$ depends on both $\theta$ and $\varphi$, the torsion vector coupling vanishes only if $P^2=F^{1-\mu}$ and $G=0$. Let us first assume that this is the case. Then there is no problem with the scalar perturbations. As in the non-minimally coupled scalar field cases discussed in \sec{sec:zero}, the theory is identical to the metric formulation if the tetrad postulate holds in the Jordan frame, and to the Palatini formulation if it holds in the Einstein frame. We have a two-field model unless $P^2=F$ (\ie $\mu=0)$. A particular case is $\mu=1$ (\ie $P=1$), when the tetrad postulate holds in the original Jordan frame and we can write $f(T+2\mu \mathring\nabla_\alpha T^\alpha, \varphi)=f(-\mathring R, \varphi)$ directly in the action (as \re{R} shows), so we get the metric formulation of $f(R,\varphi)$ theory \cite{Sotiriou:2008}. If $\mu=0$, the tetrad postulate holds in the Einstein frame, and the theory reduces to a single-field model that is identical to the Palatini formulation of $f(R,\varphi)$ theory \cite{ShahidSaless:1987, Sotiriou:2006, Sotiriou:2008, Olmo:2011}. In particular, for $f(T)=-T+2\a T^2$, the only effect of the $T^2$ term during slow-roll inflation is to suppress the tensor-to-scalar ratio \cite{Enckell:2018a}. However, it is not possible to use this mechanism to bring the predictions of the Higgs inflation model based on the dominance of $G'{}^2$ in the kinetic term (discussed at the end of \sec{sec:Higgs}) into agreement with observations. The reason is that if the original action is non-linear in $T$, and $G'\neq0$, then the coupling to the torsion vector is necessarily non-zero in the Einstein frame final action, leading to problems with perturbations.

Let us then consider the case when we have $P^2=F$, but $\mu\neq0$ or $G'\neq0$. We then have  a single field model with a derivative coupling to the torsion vector, as the variation with respect to $\theta$ gives an algebraic equation that determines $\theta$ as a function of $\varphi$, $g^{\a\b}\pat_\a\varphi\pat_\b\varphi$ and $T^\a\pat_\a\varphi$. The resulting dependence on the torsion vector is, in general, rather complicated, and it is not straightforward to see whether linear scalar perturbations can be non-zero.

\subsubsection{The special case $f(T+2\mu\mathring\nabla_\alpha T^\alpha)$}

Let us now consider the special case when there is no scalar field to begin with, so the gravity part of the action is proportional to $f(T+2\mu\mathring\nabla_\a T^\a)$ \cite{Ferraro:2006jd, Ferraro:2008ey, Bengochea:2008gz, Cai:2015emx, Bahamonde:2015zma, Bahamonde:2015hza}. This case corresponds to \re{ftbaction-scalarfield} with $K=G=0$, and no dependence on $\varphi$ in $F$, $P$ and the matter action. The only case with non-zero scalar perturbations is the one where the derivative coupling to the torsion vector vanishes, $P^2=F^{1-\mu}$. The Einstein frame action then reads
\begin{dmath}[label={ftminaction2}]
  S = \int \dd[4]{x} e \left[ \ha \mathring R - \frac{3}{4} \mu^2 \frac{F'{}^2}{F^2} g^{\a\b} \pat_\a \theta \pat_\b \theta + \frac{ \theta F + f }{ 2 F^2 } \right] + S_{\textrm{(m)}}[\Psi, F^{-1/2} e^A{}_\a, \mathring\omega^A{}_{\a B}(F)] \ ,
\end{dmath}
where prime now denotes derivative with respect to $\theta$.

If $\mu=1$, we have $P'=0$, and the tetrad postulate holds in the original Jordan frame used in \re{ftbaction}. Then the action \re{ftminaction2} corresponds to $f(\mathring R)$ theory in the metric formalism, with its extra dynamical scalar field \cite{Sotiriou:2008}. Note that any value $\mu\neq0$ will lead to a viable theory with perturbations, but the theory agrees with the metric $f(R)$ case only if $\mu=1$.

If $\mu=0$, the tetrad postulate holds in the Einstein frame, $P^2=F$. Then the scalar field does not have a kinetic term, and its equation of motion gives a constant value of $\theta$. The gravitational part reduces to the Ricci scalar plus a cosmological constant, and the non-trivial form of $f$ just contributes to the value of the latter. This is identical to the result for $f(R)$ theory in the Palatini formalism. In this case there is no extra scalar degree of freedom (and hence no scalar perturbations unless supported by the matter in $S_{\textrm{(m)}}$), as is well known \cite{Stephenson:1958, Stephenson:1959, ShahidSaless:1987, Sotiriou:2006, Sotiriou:2008, Olmo:2011}.

\section{Discussion} \label{sec:disc}

\para{The number of degrees of freedom and stability.}

Consider the non-minimally coupled theories that have no problem with linear perturbations. The simplest possibilities are those where the tetrad postulate is satisfied in the original Jordan frame or in the Einstein frame. The theory then reduces to the metric or the Palatini formulation, respectively. This shows that the equivalence between the teleparallel and the metric (or the Palatini) formulation holds for a wider class of theories than the minimally coupled Einstein--Hilbert case. However, with a different choice for the tetrad postulate function, the teleparallel theory has new features, and does not reduce to the metric nor the Palatini theory. In any case, the theories have the two usual massless propagating gravitational degrees of freedom of the metric plus the scalar field degree of freedom.

As for the theories where there are no linear scalar perturbations, it is possible they suffer from linearisation instability, so that solutions of the linearly perturbed equations are not a linearisation of the solutions of the full equations. Often linearisation instability goes in the other direction, with the linear equations missing non-linear constraints. In the present case, we would instead have the situation that there are non-linear perturbations that the linear equations miss, which could point to a strong coupling problem around the FRW background \cite{Izumi:2012qj, Golovnev:2018wbh, Koivisto:2018loq, Jimenez:2019tkx, Jimenez:2019b}.
 
For the $f(T)$ theory there are different results in the literature regarding the number of propagating degrees of freedom \cite{Chen:2010va, Li:2011rn, Izumi:2012qj, Chen:2014qtl, Ferraro:2018axk, Ferraro:2018tpu, Golovnev:2018wbh, Koivisto:2018loq}. We find that there are no cases where there are scalar perturbations around the FRW background. One subcase is equivalent to the Palatini formulation of $f(R)$ theory, where the absence of extra degrees of freedom is known to hold at the non-linear level \cite{Stephenson:1958, Stephenson:1959, ShahidSaless:1987, Sotiriou:2006, Sotiriou:2008, Olmo:2011}. However, other cases may suffer from linearisation instability or strong coupling. In the $f(T+2\mu\mathring\nabla_\a T^\a)$ theory (with $\mu\neq0$), the cases where there are scalar perturbations reduce to the usual Einstein--Hilbert action plus a minimally coupled scalar field with a positive kinetic term. This means there are three propagating modes (the two usual massless gravitons, plus one scalar), as in the metric $f(R)$ theory. Whether the scalar is massless, massive or tachyonic depends on the effective potential $U(\theta)=(\theta f'-f)/(2 f'{}^2)$ appearing in \re{ftminaction1}.

We would expect that from the point of view of the quantum theory, both non-minimal coupling and non-linear torsion scalar terms have to be included in the action (even if we restrict to only dimension 4 terms), as in the metric and the Palatini case. Even the simplest terms can lead to complicated phenomenology, as in the metric case \cite{Barbon:2015, Salvio:2015kka, Salvio:2017oyf, Kaneda:2015jma, Calmet:2016fsr, Wang:2017fuy, Ema:2017rqn, Pi:2017gih, He:2018gyf, Gorbunov:2018llf, Ghilencea:2018rqg, Wang:2018kly, Gundhi:2018wyz, Karam:2018, Kubo:2018, Enckell:2018c, Ema:2019, Canko:2019mud}. In the teleparallel case, keeping to parity-conserving terms, we could couple the scalar field separately to the three different dimension 2 scalars formed from the torsion tensor that appear in \re{QT}. This would be a generalisation of new general relativity, where these terms appear with constant coefficients  \cite{PhysRevD.19.3524, PhysRevD.24.3312, Jimenez:2019tkx}. We could also include more complicated covariant derivatives of the torsion tensor than $\mathring\nabla_\a T^\a$. If we do not require parity to be conserved, new terms quadratic in torsion can be constructed, and we could also include a derivative coupling of the scalar field to the axial torsion vector $\hat T^\a=\frac{1}{6}\epsilon^{\a\b\c\d}T_{\b\c\d}$ (coupling to a pseudoscalar field would give a parity-conserving term).

How stable is the result that there are no scalar perturbations to adding new torsion terms? If the scalar field does not generate scalar perturbations to begin with, the higher order terms we have considered do not change the situation. If there is originally no problem with the scalar perturbations, the situation is less clear. If the original coupling to the torsion vector is zero ($G=0$), the situation corresponds to the Palatini formulation, and we retain a single-field theory with scalar perturbations when adding higher order terms in $T$ (possibly coupled to the scalar field), though not if we add terms that are non-linear in $T+2\mu\mathring\nabla_\a T^\a$ with $\mu\neq0$. Likewise, if we consider the metric-equivalent case with $P=1$ and $G=F$, and add higher order terms in $T+2\mathring\nabla_\a T^\a$, the scalar perturbations remain, although we get a two-field theory, as is well known in the metric formulation \cite{Salvio:2015kka, Kaneda:2015jma, Calmet:2016fsr, Wang:2017fuy, Ema:2017rqn, Pi:2017gih, He:2018gyf, Gorbunov:2018llf, Ghilencea:2018rqg, Wang:2018kly, Gundhi:2018wyz, Karam:2018, Kubo:2018, Enckell:2018c, Canko:2019mud}. In contrast, if $F'\neq G'\neq0$, adding terms non-linear in $T$ or $T+2\mu\mathring\nabla_\a T^\a$ generates a coupling to the torsion vector. We then have either a two-field model or a single-field model with complicated dependence on the torsion vector, and the situation with scalar perturbations is not clear. Thus, taking into consideration stability to higher order terms, it remains an open question whether the only teleparallel theories that have no problem with perturbations are those that are equivalent to metric or Palatini theories.

The strong coupling problems have been discussed in the literature \cite{Izumi:2012qj, Golovnev:2018wbh, Koivisto:2018loq, Jimenez:2019tkx, Jimenez:2019b}, although many papers have missed the issue, sometimes by neglecting the antisymmetric part of the tetrad equations or not considering how the perturbations of the extra degrees of freedom behave. It is not clear for which backgrounds (other than spatially flat FRW) the linear scalar perturbations are zero. The general answer to this question requires a non-perturbative Hamiltonian analysis \cite{Li:2011rn, Ferraro:2018tpu, Ferraro:2018axk}.

\section{Conclusions} \label{sec:conc}

\para{Results and open issues.}

We have considered gravity in the teleparallel formulation, coupled to a scalar field $\varphi$. A non-minimally coupled scalar field can potentially distinguish between different formulations of gravity. Phrased differently, different formulations of general relativity that are equivalent for the Einstein--Hilbert action with minimally coupled matter can potentially provide different phenomenology when scalar fields are present. As we know that there is a Higgs field, this is not optional. In addition to the coupling $F(\varphi)$ to the torsion scalar and $G(\varphi)$ to the torsion vector, we have also included the coupling function $P(\varphi)$ that takes into account that the tetrad postulate can be taken to hold in different conformal frames.

Transforming to the Einstein frame, we find that the scalar field sources linear scalar perturbations only if the total coupling to the torsion vector due to the functions $F$, $G$ and $P$ vanishes. If this is the case, the theory can be equivalent to the metric or the Palatini formulation, or can have new kind of behaviour, depending on the relation between the three functions. In particular, for Higgs inflation, restricting to dimension 4 terms and the tree-level potential, we retain the successful prediction for the spectral index and can raise the tensor-to-scalar ratio up to any value.

Theories based on the $f(T)$ Lagrangian are just theories that are linear in $T$ but have a scalar field, written in different coordinates in field space, so our results apply to them as well. In none of them can the gravity sector alone support linear scalar perturbations. However, theories based on the Lagrangian $f(T+2\mu\mathring\nabla_\a T^\a)$, with a constant $\mu\neq0$, allow linear scalar perturbations, for suitable choices of the relation between $f$ and $P$. For $\mu=1$, these theories reduce to the metric formulation of $f(R)$ theories, as is well known, but there are also other cases.

Considering $f(T+2\mu\mathring\nabla_\a T^\a, \varphi)$ theories, where we have both a non-linear gravitational action and a non-minimally coupled scalar field, allows us to consider the stability of the results to adding more complicated torsion terms. If linear scalar perturbations are disallowed by the lower order terms already, such terms do not change the situation. In the opposite case where linear scalar perturbations are allowed by the lowest order action, the new terms cannot remove the linear scalar perturbations when the teleparallel formulation is equivalent to either the metric or the Palatini formulation. In other cases the issue remains open. It also remains an open problem how the results generalise to teleparallel theories with a more complicated gravitational sector that includes invariants other than the torsion scalar and the gradient of the torsion vector.

\acknowledgments

We thank Alexey Golovnev and Tomi Koivisto for helpful discussions and correspondence.

\bibliographystyle{JHEP}
\bibliography{tele}

\end{document}